\documentclass[preprint,showpacs,epsfig,prb]{revtex4}
\usepackage{graphicx}
\usepackage{epsfig}

\begin{document}
\title{Dispersion of the dielectric function of a charge-transfer insulator}
\author{R. O. Kuzian}
\affiliation{Institute for Problems of Materials Sciences,\\
Krzhizhanovskogo 3, 03180 Kiev, Ukraine}
\author{R. Hayn}
\affiliation{Leibniz-Institut f\"ur Festk\"orper- und 
Werkstoffforschung Dresden,\\
P.O.Box 270016, D-01171 Dresden, Germany}
\affiliation{Laboratoire Mat\'eriaux et Micro\'electronique de Provence, \\
49, rue Joliot-Curie, IRPHE, F-13384 Marseille Cedex 13, France}
\author{A.F. Barabanov}
\affiliation{Institute for High Pressure Physics, \\
142190 Troitsk, Moscow Region, Russia}
\date{\today}

\begin{abstract}
We study the problem of dielectric response in the strong coupling
regime of a charge transfer insulator. The frequency and wave number
dependence of the dielectric function $\varepsilon ({\bf q},\omega )$ and its
inverse $\varepsilon ^{-1}({\bf q},\omega )$ is the main object of
consideration. We show that the problem, in general, cannot be reduced to
a calculation within the Hubbard model, which takes into account only a 
restricted number of electronic states near the Fermi energy. The contribution
of the rest of the system to the longitudinal response (i.e. to $\varepsilon
^{-1}({\bf q},\omega )$) is essential for the whole frequency range. With
the use of the spectral representation of the two-particle Green's function 
we show that the problem may be divided into two parts: into the contributions
of the weakly 
correlated and the Hubbard subsystems. For the latter we propose an approach
that starts from the  
correlated paramagnetic ground state with strong antiferromagnetic
fluctuations. We obtain a set of coupled equations of motion for the
two-particle 
Green's function that may be solved by means of the projection technique. The
solution is expressed by a two particle basis that includes the excitonic
states with electron and hole separated at various distances.
We apply our method to the multiband Hubbard (Emery) model that describes
layered cuprates. We show that strongly dispersive branches exist in
the excitonic spectrum of the 'minimal' Emery model ($1/U_d=U_p=t_{pp}=0$) and
consider the dependence of the spectrum  on finite oxygen hopping $t_{pp}$ and
on-site repulsion $U_p$. The relationship of our  calculations to electron
energy loss spectroscopy is discussed. 
\end{abstract}


\maketitle

\section{Introduction}

The importance of many body effects for the description of dielectric response
of insulating solids is generally accepted. In the neighborhood of the
fundamental absorption threshold bound exciton lines and continuum excitons
drastically change the spectrum. Recently it has been realized that the
excitons in charge-transfer insulators (CTI) possess unusual features
connected with the strongly correlated character of the antiferromagnetic (AFM)
ground state in these compounds. The most pronounced peculiarity consists in
the existence of exciton branches with substantially larger dispersion
compared with the one-particle excitations. \cite{Wang96,Fink01} This
behavior has a formal analogy with that of the Frenkel exciton \cite{Frenkel}
that acquires a finite effective mass although both, electron and hole, have
infinite masses, \cite{Egri} but in CTI it has a completely different origin.

The experimental technique suitable for the observation of the exciton
dispersion is the electron energy loss spectroscopy (EELS).\cite{Pines} What
is actually measured in transmission EELS experiments is the partial cross
section\cite{Fink89,Schnat79} that may be decomposed into an
amplitude factor and a dynamic structure factor%
$$
\frac{d^2\sigma }{d\Omega dE}=\frac 4{(a_0)^2q^4}S({\bf q},\omega ). 
$$
The dynamic structure factor characterizes the linear response of the whole
electronic system on {\em longitudinal} electric fields with the momentum $
{\bf q}$ and frequency $\omega $ ( the ionic contribution may be neglected
for the considered frequency range of the order of several eV). Pronounced
peaks in $S({\bf q},\omega )$ are called excitons. \cite{remark} 
They may correspond to
discrete lines in the excitation spectrum of the solid or to resonances in the
continuum part of the spectrum.

For the theoretical description of excitonic features in conventional
semiconductors and insulators the following scheme is used (The key
references are Refs.\ \onlinecite{Egri,Knox,HankeShm,AbInit98,Prokof01}). 
First, the quasiparticle excitation spectrum is found
empirically \cite{Egri,Knox,Prokof01} or from first principles. 
\cite{HankeShm,AbInit98} It is essential that the spectrum of the 
$N$-electron system consists mainly of a continuum of electron-hole pairs 
whose electron or hole quasiparticle excitations are close to eigenstates of
the $(N+1)$- or $(N-1)$-electron system, respectively, with definite 
quasimomenta {\em and} energy. \cite{ShRice}
The quasiparticle spectrum is usually obtained from the 
self-consistent-field (SCF) approach.
Next, the electron-hole interaction is taken into account. Then the problem
for two quasi-particles interacting via the medium is solved. It is crucial
that the ground state may be viewed as an occupied valence band that is
separated from the first excited state by an energy gap.

For the CTI the above scheme should be revised beginning from the first
step. The CTI has an odd number of electrons per formula unit. That is why the
SCF calculations usually give a {\em metallic} ground state for the CTI and a 
gapless excitation spectrum. More elaborate SCF methods like the unrestricted
Hartree-Fock method or its modern version LDA+U predict an AFM long-range
order in the 
ground state and a gap in the excitation spectrum. Nevertheless, the nature of
excitations and the ground state fluctuations are not caught by these
approaches. Let us recall that any CTI remains insulating and shows excitonic
features in optics and EELS spectra above the N\'eel temperature, i.e.\ in the 
{\em paramagnetic} state (with short-range AFM correlations). For strongly
correlated systems the electron-electron interaction should be taken into
account beyond the mean-field level. This is possible within the framework
of Hubbard-like models in the restricted subspace of orbitals close to the 
Fermi level.

The description of exciton physics in CTI is possible within the framework
of one-band models, \cite{Penc98,Tsut99,Essler01,Eder02}
but more detailed and realistic information may only be obtained from the 
multiband Hubbard (Emery) model that explicitly includes ligand ion degrees
of freedom. For quasi-one-dimensional cuprates the Emery model was
considered in Refs.\ \onlinecite{Richt00,Hubsch01,Atzkern01,Moskv03}. 
The phenomenon of spin-charge separation that is characteristic to
one dimension introduces a specific physics into the exciton formation. The 
mobility of a single electron or hole in 1D is not suppressed by spin
correlations  and the 
exciton dispersion is comparable with the one-particle dispersion.

In quasi-two-dimensional cuprates the situation is different. Due to AFM
correlations the bandwidth of the one-particle motion is of the order of the 
AFM exchange integral $J$ which is considerably smaller than the bare hopping
$t$, and the 
exciton dispersion is of the order of $t$. The authors of Ref.\ 
\onlinecite{ZhNg}
proposed a qualitative physical explanation of the large exciton dispersion in
layered cuprates: the propagation of an electron-hole pair does not disturb
the AFM background, in contrast to the motion of a single electron or hole.
Unfortunately, the calculations of Ref.\ \onlinecite{ZhNg} do {\em not} 
support this idea since they give no dispersion in the absence of oxygen
on-site Coulomb repulsion $U_p$ and direct O-O hopping $t_{pp}$, which was
also pointed out in Ref.\ \onlinecite{Richt00}. One should expect a
qualitative description of the EELS spectra \cite{Wang96,Fink01} already for
the 'minimal' version of the Emery model ($1/U_d=U_p=t_{pp}=0$) which can be
refined by taking into account additional parameters like e.g.\ $t_{pp}$ or
$U_p$. On the contrary, the authors of Ref.\ \onlinecite{ZhNg} had to assume
quite unrealistic parameter values to fit the experimental spectra. 

In this paper we outline an approach to calculate the longitudinal and the 
transverse dielectric response (Section II) for CTI within the framework
of the 
multiband Hubbard model (Section III). Introducing an analogy to Wannier's
excitonic representation \cite{Wannier} we obtain a set of coupled equations 
of motion for the two-particle Green's function that may be solved by means of
the projection technique \cite{Fulde} (Section IV). It is substantial that the
method allows a systematic improvement of approximations. In order to retain
only the essential properties of CTI we first consider the 'minimal' version
of the 
Emery model describing the CuO$_2$ plane of high-T$_c$
superconductors and their parent compounds. The model reflects the main
features of CTI: the existence of two kinds of states - strongly correlated
'copper' states with the prohibition of double occupancy and uncorrelated
'oxygen' states; the undoped plane has one hole per unit cell, and is a 
quantum two-dimensional AFM insulator possessing no long-range order at any
finite temperature; the charge excitation corresponds to the transfer of a hole
from copper to the adjacent oxygen site. In Section V,
we discuss the solution of our equations within the small exciton basis and
its dependence on various parameters of the extended Emery model. We obtain an
appreciable exciton dispersion and a rough, qualitative agreement with the
experimental spectra already within the 'minimal' version of the Emery
model. 
%
%
%

\section{Density-density correlation function}

The dynamic structure factor is related to the density-density correlation
function

\begin{equation}
\label{Sqw}S({\bf q},\omega )\equiv \frac 1{2\pi N}\int_{-\infty }^\infty dt
{\rm e}^{-\imath \omega t}\left\langle \hat n_{{\bf q}}(t)\hat n_{-{\bf q}
}(0)\right\rangle =\frac 1\pi \frac 1{\exp (-\beta \omega )-1}{\rm Im}N_{
{\bf q}}(\omega ), 
\end{equation}
where 
\begin{equation}
\label{nq}\hat n_{{\bf q}}=\frac 1{\sqrt{N}}\sum_{{\bf r},s}\exp (-\imath 
{\bf qr})\left( a_{{\bf r},s}^{\dagger }a_{{\bf r},s}-\left\langle a_{{\bf r}
,s}^{\dagger }a_{{\bf r},s}\right\rangle \right)  
\end{equation}
is the electronic density operator in the localized basis, the summation runs
over all lattice sites ${\bf r}$ and orbital sorts $s$ ; $\left\langle
\ldots \right\rangle $ means the thermodynamic average. For 
$\beta \omega \gg 1$ 
we have
$$
S({\bf q},\omega )\approx -\frac 1\pi {\rm Im}N({\bf q},\omega ), 
$$
where 
\begin{equation}
\label{nqnq}N({\bf q},\omega )\equiv \left\langle \left\langle \hat n_{{\bf q
}}|\hat n_{-{\bf q}}\right\rangle \right\rangle =-i\int_0^\infty dt{\rm e}
^{-\imath \omega t}\left\langle \left[ \hat n_{{\bf q}}(t),\hat n_{-{\bf q}
}(0)\right] \right\rangle 
\end{equation}
is the retarded Green's function that defines the inverse dielectric
function 
\begin{equation}
\label{inveps}\varepsilon ^{-1}({\bf q},\omega )=1+\frac{4\pi e^2}{v_cq^2}N(
{\bf q},\omega ), 
\end{equation}
with $v_c$ being the volume of the unit cell, and $e$ is the electronic 
charge. We
neglect here the local field effects (coupling with the Fourier components
with ${\bf q}^{\prime }={\bf q}+{\bf G}$, ${\bf G}$ being a reciprocal lattice
vector). The function $N({\bf q},\omega )$ describes the response to the {\em 
unscreened} external potential. It requests the account of the macroscopic
electric field, i.e. the long-range part of the Coulomb interaction. The latter
is responsible for the splitting into longitudinal and transverse excitons
for small wave number which is analogous to the splitting into longitudinal
and transverse optical phonons. \cite{Knox} The response to the total, 
{\em screened} potential is given by\cite{PinesNoz} 
\begin{equation}
\label{Ns}N_s({\bf q},\omega )=\varepsilon ({\bf q},\omega )N({\bf q},\omega
), 
\end{equation}
then 
\begin{equation}
\label{eps2Ns}\varepsilon ({\bf q},\omega )=1-\frac{4\pi e^2}{v_cq^2}N_s(
{\bf q},\omega ). 
\end{equation}
In the diagrammatic language the linear response to the total field may be
expressed by the polarization operator where only irreducible graphs (which do
not contain the contribution of the macroscopic electric field) should be taken
into account. \cite{AmKo,AGD} The random-phase approximation results in  
\begin{equation}
\label{RPA}N({\bf q},\omega )=\frac{N_s({\bf q},\omega )}{1-\frac{4\pi e^2}{
v_cq^2}N_s({\bf q},\omega )} 
\end{equation}
which follows from (\ref{Ns}) and (\ref{inveps}) and which is {\em exact} 
for $
q\rightarrow 0$, as it was shown in Ref.\ \onlinecite{AmKo}.

The construction of Hubbard-like models for strongly correlated systems has
an input from LDA band structure calculations where the screening of
the long-range part of the Coulomb interaction is already taken into account. 
The
Hubbard terms arise from the short-range residual interaction. Thus, the 
density
response function $N_H({\bf q},\omega )$ calculated within the Hubbard model
is an approximation to $N_s({\bf q},\omega )$. \cite{Penc98} In other words,
it describes the motion of transverse (or 'mechanical' by Agranovich's \cite
{Agran} terminology) excitons.

Using the spectral representation we may write 
\begin{equation}
\label{Ns2NH}N_s({\bf q},z)=\int_0^\infty \left[ -\frac 1\pi {\rm Im}N_s(
{\bf q},\omega ^{\prime })\right] \frac{2\omega ^{\prime }d\omega ^{\prime }
}{z^2-\omega ^{\prime 2}}=\int_0^{\omega _0}+\int_{\omega _0}^\infty =N_H(
{\bf q},z)+N_\infty ({\bf q},z). 
\end{equation}
Here we bear in mind that the Hubbard model contributes to transitions in
the low frequency region $\omega <\omega _0$ with $\omega_0$ of the order of
the bandwidth, and the electrons of the rest
of the solid are excited only at higher energies. In zero approximation we
may assume that in the frequency region $\omega > \omega _0$ the electronic
polarization of the rest of the solid  follows immediately the external field
$$
N_\infty ({\bf q},z)\approx N_\infty ({\bf q},0). 
$$
In other words, the Hubbard model is embedded into the medium with
dielectric permeability
$$
\varepsilon _\infty ({\bf q})=1-\frac{4\pi e^2}{v_cq^2}N_\infty ({\bf q},0). 
$$
In fact, $\varepsilon _\infty $ may has its own dispersion and may be quite 
anisotropic for layered or quasi-one-dimensional compound. In principle, it
should be taken from e.g.\ LDA calculations (we have assumed that the rest of
solid is uncorrelated) or from experiment. It is obvious that the peak 
positions of the loss function 
\begin{equation}
\label{lf}L({\bf q},\omega )\equiv -{\rm Im}\left[ \varepsilon ^{-1}({\bf q}
,\omega )\right] 
\end{equation}
and their intensity strongly depend on the value of $\varepsilon _\infty (
{\bf q})$. Usually one neglects the ${\bf q}$-dependence and the anisotropy 
of $\varepsilon _\infty $ , but it is a crude approximation, as well as
another one which assumes $\varepsilon ({\bf q},0)={\rm const}$ (see the
discussion in Sec. V). For a quantitative description of EELS experiments
the detailed knowledge of $\varepsilon _\infty ({\bf q})$ is necessary.
Then the total dielectric function is 
\begin{equation}
\label{epsap}\varepsilon ({\bf q},\omega )=\varepsilon _\infty -\frac{4\pi
e^2}{v_cq^2}N_H({\bf q},\omega )\equiv \varepsilon _\infty \varepsilon _H 
\end{equation}
and its inverse
$$
\varepsilon ^{-1}({\bf q},\omega )=\varepsilon _\infty ^{-1}\varepsilon
_H^{-1}. 
$$
For the dielectric function of the Hubbard model $\varepsilon _H$ the usual 
sum rule holds
$$
\int_0^\infty \omega {\rm Im} \varepsilon _H({\bf q},\omega )d\omega
=-\int_0^\infty \omega {\rm Im}\varepsilon _H^{-1}({\bf q},\omega )d\omega =
\frac \pi 2\frac{4\pi e^2}{\varepsilon _\infty v_cq^2}\left\langle \left[
\left[ \hat n_{{\bf q}},\hat H_H\right] ,\hat n_{-{\bf q}}\right]
\right\rangle , 
$$
where $\hat H_H$ is the Hubbard model Hamiltonian and the operator 
$\hat n_{{\bf q}}$ acts in the subspace of orbitals which enter into 
$\hat H_H$ (i.e.\ the summation
over $s$ in Eq.\ (\ref{nq}) is restricted to these orbitals). Then for the
total dielectric function approximated by Eq.\ (\ref{epsap}) we have 
\begin{equation}
\label{srutot}-\int_0^\infty \omega {\rm Im}\varepsilon ^{-1}({\bf q},\omega
)d\omega =\frac \pi 2\frac{4\pi e^2}{\varepsilon _\infty ^2v_cq^2}
\left\langle \left[ \left[ \hat n_{{\bf q}},\hat H_H\right] ,\hat n_{-{\bf q}
}\right] \right\rangle =\frac 1{\varepsilon _\infty ^2}\int_0^\infty \omega 
{\rm Im}\varepsilon ({\bf q},\omega )d\omega . 
\end{equation}
The factor $1/\varepsilon _\infty ^2$ arises due to the neglect of the 
frequency dependence of $\varepsilon _\infty $.

\section{Model Hamiltonian and density operator}

As we have mentioned in the Introduction, we consider the 'minimal' Emery
model that exhibits the essential properties of layered cuprates ($
1/U_d=U_p=t_{pp}=0$). Then the total Hamiltonian in hole notation reads 
\begin{equation}
\label{Htot}\hat H_H=\hat H_0+\hat V, 
\end{equation}
where
$$
\hat H_0=\Delta \sum_{{\bf r},\gamma }\bar p_{{\bf r},\gamma }^{\dagger }
\bar p_{{\bf r},\gamma },\ \hat V=t\sum_{{\bf R,}\alpha ,\gamma }\left( \bar 
p_{{\bf R+a}_\alpha ,\gamma }^{\dagger }\bar Z_{{\bf R}}^{0\gamma }+\bar Z_{
{\bf R}}^{\gamma 0}\bar p_{{\bf R+a}_\alpha ,\gamma }\right) , 
$$
and where the Fermi 
operator $\bar p_{{\bf r},\gamma }$ annihilates a hole at site ${\bf r}$ of
the oxygen sublattice with spin projection index $\gamma $, the Hubbard 
projection 
operator $\bar Z_{{\bf R}}^{0\gamma }=\bar d_{{\bf R}\gamma}(1-n_{{\bf R} \bar
\gamma} )$ annihilates a hole with spin index 
$\gamma$ on a  
{\em singly occupied} copper site, where $\bar d_{{\bf R}\gamma}$ is the
corresponding Fermi operator. The  double occupancy of copper sites is 
thus excluded from (\ref{Htot}). $\hat H_0$ includes the on-site energies 
($\Delta =\epsilon _p-\epsilon _d,$ $\epsilon _d$ is 
taken as zero of energy), $\hat V$ is the $p$-$d$ hybridization, 
$\alpha =x,-x,y,-y$ characterizes the direction of 
a nearest-neighbor vector ${\bf a}$; the phase factors in $
\hat V$ are absorbed into the definition of the 
operators $\bar p_{{\bf r},\sigma
},\bar Z_{{\bf R}}^{0\gamma }$ , they do not change the exciton dispersion.

Taking the limit $U_d/\Delta \rightarrow \infty $ considerably simplifies
the consideration and is a good approximation for weakly doped compounds in
a wide range of values $U_d/\Delta >2$ (see e.g. Ref.\ \onlinecite{Stef97}). 
The
conditions $U_p=t_{pp}=0$ are introduced for simplicity and may be easily
relaxed (see Appendix), then the Hubbard on-site term for $p$-orbitals and the
direct O-O hopping of the form $\hat t_{pp}=-t_{pp}\sum_{\left\langle
i,j\right\rangle ,\gamma }\bar p_{{\bf r}_i,\gamma }^{\dagger }\bar p_{{\bf r
}_j,\gamma }$ is added to $\hat H_0$ .

It is well known that the Hamiltonian (\ref{Htot}) has an insulating ground
state and  is equivalent to the nearest-neighbor AFM
Heisenberg model in the low-energy region. It means that charge 
fluctuations in $\hat V$ are strongly
suppressed and that holes are localized. This fact becomes more apparent if we
make a canonical transformation of operators of the form 
$$
\hat A_{eff}=\exp (-\hat S)\hat A\exp (\hat S)=\hat A+\left[ \hat A,\hat S
\right] +\cdots \; ,  
$$
where%
$$
\hat S=-\frac t\Delta \sum_{{\bf R,}\alpha {\bf =}\pm x,\pm y,\gamma }\left(
p_{{\bf R+a}_\alpha ,\gamma }^{\dagger }Z_{{\bf R}}^{0\gamma }-Z_{{\bf R}
}^{\gamma 0}p_{{\bf R+a}_\alpha ,\gamma }\right) . 
$$
Then $\hat H_H$ becomes 
\newpage
\begin{eqnarray}
\label{Heff}\hat H_{eff} &\approx & \hat H_0-4\tau \sum_{{\bf R},\gamma} 
Z_{{\bf R}}^{\gamma \gamma }
+\tau \sum_{{\bf R},\alpha 1,\alpha 2,\gamma }p_{{\bf R}
+{\bf a}_{\alpha 1},\gamma _1}^{\dagger }p_{{\bf R}
+{\bf a}_{\alpha 2},\gamma
_2}\left( Z_{{\bf R}}^{00}\delta _{\gamma _1\gamma _2}+Z_{{\bf R}}^{\gamma
_2,\gamma _1}\right) 
\nonumber \\
&&
-\tau \sum_{{\bf R},\alpha ,\gamma _1}Z_{{\bf R}+{\bf g}
_\alpha }^{\gamma _10}Z_{{\bf R}}^{0\gamma _1}+\hat J_s \; , 
\end{eqnarray}
(see also Ref.\ \onlinecite{Kuz98} for the notation). 
Here $p$ and $Z$ mean transformed operators, $\hat J_s$ is the AFM 
copper-copper
superexchange interaction, ${\bf g}$ points to neighboring copper sites.
Strictly speaking, the Hamiltonian (\ref{Heff}) is obtained under the 
condition $t/\Delta \ll 1$, and its parameters are $\tau =t^2/\Delta $ and the
AFM exchange $J\propto t^4/\Delta ^3$. Nevertheless, it may be applied in a 
wider range $t/\Delta <1$ with renormalized values of $\tau$ and $J$.

The advantage of using the effective Hamiltonian (\ref{Heff}) instead of the 
bare one (\ref{Htot}) consists in excluding irrelevant zero-point charge
fluctuations. Then the coupling of carriers with spin fluctuations that
governs the low-energy physics of CTI becomes apparent. It is essential that
the effective Hamiltonian (\ref{Heff}) does not contain transitions between 
$p$- and $Z$-states, in other words, it never creates a  particle-hole state
out of the dielectric state. In this sense it resembles the starting 
Hamiltonian for the 
transverse exciton motion in conventional insulators. This allows to
introduce an analog of Wannier's excitonic representation for the description 
of the electron-hole pair dynamics.

The bare density operator 
\begin{equation}
\label{nqEm}\hat n_{{\bf q}}=\frac 1{\sqrt{N}}\left[ \sum_{{\bf R},\sigma
}\exp (-\imath {\bf qR})\left( \bar Z_{{\bf R}}^{\sigma \sigma
}-\left\langle \bar Z_{{\bf R}}^{\sigma \sigma }\right\rangle \right) +\sum_{
{\bf s}}\exp (-\imath {\bf qs})\left( \bar p_{{\bf s},\sigma }^{\dagger }
\bar p_{{\bf s},\sigma }-\left\langle \bar p_{{\bf s},\sigma }^{\dagger }
\bar p_{{\bf s},\sigma }\right\rangle \right) \right] , 
\end{equation}
(${\bf s}$ runs over O sublattice) is transformed to 
\begin{equation}
\label{nqtr}\hat n_{{\bf q}}=\frac 1{\sqrt{N}}\left\{ \sum_{{\bf R},\sigma
}\exp (-\imath {\bf qR})\left[ Z_{{\bf R}}^{\sigma \sigma }+\frac t\Delta
\sum_{\alpha {\bf =}\pm x,\pm y}\left( p_{{\bf R+a}_\alpha ,\sigma
}^{\dagger }Z_{{\bf R}}^{0\sigma }+Z_{{\bf R}}^{\sigma 0}p_{{\bf R+a}_\alpha
,\sigma }\right) \right] +\right. 
\end{equation}
$$
\left. \sum_{{\bf R,}\alpha {\bf =}+x,+y}\exp \left[ -\imath {\bf q(R+a)}
\right] \left[ p_{{\bf R+a}_\alpha ,\sigma }^{\dagger }p_{{\bf R+a}_\alpha
,\sigma }-\frac t\Delta \left[ \left( Z_{{\bf R}}^{\sigma 0}+Z_{{\bf R+2a}
_\alpha }^{\sigma 0}\right) p_{{\bf R+a}_\alpha ,\sigma }+h.c.\right]
\right] \right\} \; . 
$$
Note that in the second line of (\ref{nqtr}) (in the sum over the 
oxygen sublattice) ${\bf a}_\alpha $ lies
only in the same cell as ${\bf R}$. Collecting the terms surrounding a Cu
site, we have
$$
\hat n_{{\bf q}}=\tilde n_{{\bf q}}+\frac 1{\sqrt{N}}\frac t\Delta \sum_{
{\bf R},\sigma }\exp (-\imath {\bf qR})\sum_{\alpha {\bf =}\pm x,\pm
y}\left( p_{{\bf R+a}_\alpha ,\sigma }^{\dagger }Z_{{\bf R}}^{0\sigma }+Z_{
{\bf R}}^{\sigma 0}p_{{\bf R+a}_\alpha ,\sigma }\right) \left( 1-\exp
(-\imath {\bf qa}_\alpha {\bf )}\right) 
$$
\begin{equation}
\label{nqpsi}=\tilde n_{{\bf q}}+\frac 1{\sqrt{N}}\frac t\Delta \sum_{{\bf R}
}\exp (-\imath {\bf qR})\sum_{\alpha {\bf =}\pm x,\pm y}\left( \psi _{{\bf 
R,a}_\alpha }^{\dagger }+\psi _{{\bf R,a}_\alpha }\right) \left( 1-\exp
(-\imath {\bf qa}_\alpha {\bf )}\right) , 
\end{equation}
where the operator 
\begin{equation}
\label{psi}\psi _{{\bf R,}\alpha }\equiv \sum_\gamma Z_{{\bf R}}^{\gamma
,0}p_{{\bf R}+{\bf a}_\alpha ,\gamma } 
\end{equation}
annihilates an electron-hole pair with minimal distance, and 
\begin{equation}
\label{nqwave}\tilde n_{{\bf q}}\equiv \frac 1{\sqrt{N}}\left[ \sum_{{\bf R}
,\sigma }\exp (-\imath {\bf qR})\left( Z_{{\bf R}}^{\sigma \sigma
}-\left\langle \bar Z_{{\bf R}}^{\sigma \sigma }\right\rangle \right) +\sum_{
{\bf s}}\exp (-\imath {\bf qs})\left( p_{{\bf s},\sigma }^{\dagger }p_{{\bf s
},\sigma }-\left\langle \bar p_{{\bf s},\sigma }^{\dagger }\bar p_{{\bf s}
,\sigma }\right\rangle \right) \right] . 
\end{equation}
As we have mentioned above, the effective Hamiltonian (\ref{Heff}) conserves
the number of particles in every band. Therefore $\tilde n_{{\bf q}}=0$ gives 
no contribution to $N_H$.
Having the operator (\ref{nqpsi}) we may proceed with the calculation of
the density-density response function (\ref{nqnq}).

\section{Electron-hole pair dynamics}

The problem of the dielectric function (\ref{epsap}) calculation is thus
reduced to the calculation of the two-particle Green's function 
\begin{equation}
\label{NH}N_H({\bf q},\omega )=\left\langle \left\langle \Phi _{{\bf q}
}+\Phi _{-{\bf q}}^{\dagger }|\Phi _{{\bf q}}^{\dagger }+\Phi _{-{\bf q}
}\right\rangle \right\rangle _\omega =\left\langle \left\langle \Phi _{{\bf q
}}|\Phi _{{\bf q}}^{\dagger }\right\rangle \right\rangle _\omega
+\left\langle \left\langle \Phi _{-{\bf q}}|\Phi _{-{\bf q}}^{\dagger
}\right\rangle \right\rangle _{-\omega } 
\end{equation}
where 
\begin{equation}
\label{dipac}\Phi _{{\bf q}}\equiv \frac t\Delta \sum_\alpha \left( 1-\exp
(-\imath {\bf qa}_\alpha {\bf )}\right) \psi _{{\bf q},\alpha },\ \psi _{
{\bf q},\alpha }=\frac 1{\sqrt{N}}\sum_{{\bf R}}\exp (-\imath {\bf qR})\psi
_{{\bf R}\alpha }, 
\end{equation}
and where we used again the conservation of particle numbers in the electron 
and the 
hole subsystem that excludes anomalous Green's functions like $\left\langle
\left\langle \Phi _{{\bf q}}|\Phi _{{\bf q}}\right\rangle \right\rangle
_\omega $.

The equation of motion 
\begin{equation}
\label{eqmo}\omega \left\langle \left\langle \Phi _{{\bf q}}|\Phi _{{\bf q}
}^{\dagger }\right\rangle \right\rangle _\omega =\left\langle \left[ \Phi _{
{\bf q}},\Phi _{{\bf q}}^{\dagger }\right] \right\rangle +\left\langle
\left\langle \left[ \Phi _{{\bf q}},\hat H_{eff}\right] |\Phi _{{\bf q}
}^{\dagger }\right\rangle \right\rangle _\omega 
\end{equation}
generates more complex operators 
\begin{equation}
\label{xi}\xi _{{\bf R},\alpha ,\beta }=\sum_{\gamma _1,\gamma _2}Z_{{\bf R}+
{\bf g}_\alpha }^{\gamma _1\gamma _2}Z_{{\bf R}}^{\gamma _2,0}p_{{\bf R}+
{\bf g}_\alpha +{\bf a}_\beta ,\gamma _1} \; ,
\end{equation}
which annihilate states with an increasing separation between
electron and hole, accompanied by spin fluctuations. The set of equations of
motion will generate states corresponding to electrons and holes that are more
and more separated and dressed by spin fluctuations. These states form a set 
similar to the 
excitonic representation for conventional insulators. \cite{Wannier,Knox,Egri}
The complication that arises in CTI consists in the strong
interaction of both electron and hole with AFM fluctuations. The effect of
this interaction leads to a strong renormalization of the one-particle
bandwidth, but it is partially canceled when electron and hole follow each
other.

The set of coupled equations (\ref{eqmo}) may be approximately solved by
means of the projection technique.
We choose an operator basis $B_{{\bf q,}i}$  and the definition of the scalar
product $\langle \left[ B_{{\bf q,}i},B_{{\bf q,}j}^{\dagger }\right]
\rangle $. Within the operator subspace spanned by this basis we are
looking for the approximate solution of the eigenvalue problem 
\begin{equation}
\label{eig}\left[ \Psi _{{\bf q}},\hat H\right] =E_{{\bf q}}\Psi _{{\bf q}}
\, , \quad
\Psi _{{\bf q}}=\sum_ic_i({\bf q})B_{{\bf q,}i} \; . 
\end{equation}
This leads, as usual for a non orthonormal basis, to the generalized eigenvalue
problem 
\begin{equation}
\label{geig}\sum_ic_i({\bf q})L_{i,j}({\bf q})=E\sum_ic_i({\bf q})S_{ij}(
{\bf q}) 
\end{equation}
where overlap and Liouvillean matrices 
\begin{equation}
\label{SandL}S_{ij}\equiv \langle \left[ B_i,B_j^{\dagger }\right] \rangle
=\langle B_i,B_j^{\dagger }\rangle \; , \quad 
L_{ij}\equiv \langle \left[ \left[ B_i,
\hat H\right] ,B_j^{\dagger }\right] \rangle =\langle [B_i,\hat H
],B_j^{\dagger }\rangle 
\end{equation}
depend only on spin-spin correlation functions for the system without
electron-hole pairs, which is equivalent to the Heisenberg antiferromagnet. 
The correlation functions may thus be calculated from the Heisenberg model.

The system (\ref{geig}) may be solved numerically. Then we find the 
eigenvectors 
\begin{equation}
\label{eigf}\Psi _{{\bf q}}^\lambda =\sum_i
c_i^\lambda ({\bf q})B_{{\bf q,}i} \;  , 
\end{equation}
where 
$\lambda $ is the number of the branch in the spectrum. In order to 
calculate $N_H({\bf q},\omega )$ within our basis, we should expand 
\begin{equation}
\label{eigexp}\Phi _{{\bf q}}=\sum_\lambda 
g^\lambda ({\bf q})\Psi _{{\bf q}}^\lambda \; . 
\end{equation}
Then one obtains 
\begin{equation}
\label{pspsgf}\left\langle \left\langle \Phi _{{\bf q}}|\Phi _{{\bf q}
}^{\dagger }\right\rangle \right\rangle _\omega =\sum_\lambda \left|
g^\lambda ({\bf q})\right| ^2\left\langle \left\langle \Psi _{{\bf q}
}^\lambda |\left( \Psi _{{\bf q}}^\lambda \right) ^{\dagger }\right\rangle
\right\rangle _\omega =\sum_\lambda \left| g^\lambda ({\bf q})\right|
^2\left( \omega -E_{{\bf q}}^\lambda \right) ^{-1}, 
\end{equation}
with 
\begin{equation}
\label{gklmbd}\left| g^\lambda ({\bf q})\right| ^2=\left( \frac t\Delta
\right) ^2\left| \sum_{i,\alpha }c_i^\lambda ({\bf q})S_{i\alpha }\left[
1-\exp (\imath {\bf qa}_\alpha {\bf )}\right] \right| ^2, 
\end{equation}
and finally 
\begin{equation}
\label{epsans}\varepsilon ({\bf q},\omega )=\varepsilon _\infty -\frac{4\pi
^2e^2}{v_cq^2}\sum_\lambda \left| g^\lambda ({\bf q})\right| ^2\left[ \left(
\omega -E_{{\bf q}}^\lambda \right) ^{-1}-\left( \omega +E_{{\bf q}}^\lambda
\right) ^{-1}\right] . 
\end{equation}
Let us note that the projection technique allows to improve the chosen 
approximation step
by step by enlarging the basis set.

\section{Results and discussion}

We have restricted ourself to the minimal basis that describes the
electron-hole pair with minimal distance. The basis contains 
operators (\ref{psi}) and (\ref{xi}) with $\beta =-\alpha $. 
Then the problem (\ref{eig}) has the
dimension $8\times 8$. The overlap and Liouvillean matrices are given in the
Appendix. Spin-spin correlation functions were taken from the spherically
symmetric treatment of an S=1/2 Heisenberg AFM model on the square lattice. 
\cite{BarBer94} For a low temperature $T=0.1J$ and a vanishing frustration
parameter $p=0.01$ they have the following values 
$\left\langle {\bf \hat S}_{
{\bf R}}\cdot {\bf \hat S}_{{\bf R+g}}\right\rangle =-0.33$, $\left\langle 
{\bf \hat S}_{{\bf R}}\cdot {\bf \hat S}_{{\bf R+g}_x+{\bf g}
_y}\right\rangle =0.20,\ \left\langle {\bf \hat S}_{{\bf R}}\cdot {\bf \hat S
}_{{\bf R+2g}}\right\rangle =0.17$. For the on-site energy difference and
$p$-$d$ hopping we took the values $\Delta =3.6\ $eV and $t=1.3$ eV, which 
are
characteristic to all cuprates.

Fig.\ 1 shows the dispersion of the imaginary part of the dielectric 
function $\varepsilon _2({\bf q},\omega )
\equiv {\rm Im}\varepsilon ({\bf q},\omega )$. 
The oxygen on-site repulsion and the O-O hopping were neglected. We see 
strongly dispersive branches both in the [110] and in the [100] direction.

As we have mentioned above, the comparison with the EELS experiment may have
only qualitative character without a detailed knowledge of the background
dielectric constant $\varepsilon _\infty ({\bf q})$. Fig.\ 2 shows the graphs
for the loss function (multiplied by $\left[ \varepsilon _\infty ({\bf 0}
)\right] ^2$ in order to have approximately the same normalization as $
\varepsilon _2({\bf q},\omega )$ according to (\ref{srutot})) under the 
assumption that \cite{ZhNg} 
\begin{equation}
\label{eps0Zh}\varepsilon ({\bf q},\omega =0)\approx \varepsilon ({\bf q=0}
,\omega =0)=4.83. 
\end{equation}
Then the value of $\varepsilon _\infty$ was obtained from Eq.\ (\ref{epsans}).
In this case the dispersion in the loss function reproduces essentially 
the dispersion in $\varepsilon ({\bf q},\omega )$. The peaks are slightly
shifted to higher energies. In fact, the assumption (\ref{eps0Zh}) implies
that the dispersion of $N_\infty ({\bf q},\omega )$ should follow the
dispersion of $N_H({\bf q},\omega )$ in such a way that 
$$
\varepsilon ({\bf q},\omega =0)=1-\frac{4\pi e^2}{v_cq^2}\left[ N_H({\bf q}
,0)+N_\infty ({\bf q},0)\right] ={\rm const} 
$$
as it follows from (\ref{Ns2NH}) and (\ref{eps2Ns}).

In general, the interplay of $N_H({\bf q},\omega )$ and $N_\infty ({\bf q}
,\omega )$ should be more complex. In order to demonstrate the strong 
dependence
on the value of $\varepsilon _\infty$, we plot in Fig.\ 3 the 
same graph assuming
a constant value $\varepsilon _\infty =2$ for all ${\bf q}$. We see a 
qualitative
difference with Fig.\ 2 and we may conclude that the dependence of the loss
function on $\varepsilon _\infty $ is nontrivial. For quasi-one dimensional
compounds a large values of $\varepsilon _\infty \sim 8$ was taken in 
Refs.\ \onlinecite{Hubsch01,Atzkern01}. This means that the rest of the solid 
strongly
screens the long-range part of the Coulomb interaction between electrons that
enter the Hubbard model. For this situation the poles of 
$\varepsilon ({\bf q},\omega )$ are very close to the poles of 
$\varepsilon ^{-1}({\bf q},\omega)$ and the shape of the loss function 
is close to the shape of $\varepsilon
_2({\bf q},\omega )$. Note also that always $\varepsilon ({\bf q}
,0)>\varepsilon _\infty $ as follows from Eq. (\ref{epsap}), (\ref{epsans})
and with $\varepsilon _\infty =8$ one will receive unrealistically large $
\varepsilon ({\bf q},0)$.

Let us now show some examples for the dependence of the dielectric function on
various parameters of the model. 
For the reasons outlined
above, from now on we consider only figures for 
$\varepsilon _2({\bf q},\omega )$. 
As it was mentioned in Section III, the model (\ref{Htot}) may
be generalized in order to include a finite Hubbard repulsion $U_p$ on the 
oxygen site and a direct oxygen-oxygen hopping $t_{pp}$.

The dependence on $t_{pp}$ is not very strong for $t_{pp}<\tau $. Let us
recall that the addition of the O-O hopping to the Hamiltonian (\ref{Heff})
within the parameter range $t_{pp}=(0.3\div 0.4)\tau $ 
is essential to describe correctly the angle-resolved photoemission 
spectroscopy (ARPES) of layered cuprates. \cite{Kuz98,Bar01}
Fig.\  4 shows $\varepsilon _2({\bf q},\omega )$ for $
t_{pp}=0.4\tau $ , and $U_p=0$. We see that the main difference to Fig.\ 1
consists in the redistribution of spectral weight between the different 
branches of the spectrum.

The spectrum demonstrates a  much stronger dependence on $U_p$. Fig.\ 5 
displays 
$\varepsilon _2({\bf q},\omega )$ for $U_p=4$ eV. We have an almost vanishing
dispersion of the lower branches. Let us pay attention to the fact that $U_p$
does not affect the  
single hole motion. Due to that reason its value is experimentally not well 
established. Our
results show that the dielectric function dispersion is very sensitive to
this parameter.

Now let us discuss a very important peculiarity of our figures, namely the
absolute position of the intensive peaks in the spectrum of $\varepsilon _2(
{\bf q},\omega )$. At the $\Gamma $ point (${\bf q}=(0,0)$) we have one peak
with the energy $E_{{\bf \Gamma }}=\Delta +4\tau \approx 5.5$ eV. Let us
estimate the edge of the electron-hole continuum. It corresponds to the
energy that is needed for the excitation of electron and hole which are
independent on each other. The
operator that annihilates such a state is
$$
\Psi _{{\bf q}}^{e-h}=e_{{\bf k+q}}h_{{\bf -k}},\ \left[ \Psi _{{\bf q}
}^{e-h},\hat H_{eff}\right] =(\epsilon _{{\bf k+q}}^e+\epsilon _{{\bf -k}
}^h)\Psi _{{\bf q}}^{e-h}, 
$$
where $\epsilon _{{\bf k}}^e=-(-4\tau -\epsilon _{{\bf k}}^{t-J})$
corresponds to the energy of a single 'electron' quasiparticle, that will be a
complex spin-polaron corresponding to the coherent motion in the so 
called $t-J$
model, which describes the $Z$ subsystem in (\ref{Heff}) in the absence of
holes (with $t^{t-J}=\tau $). $\epsilon _{{\bf k}}^h=\Delta -\epsilon _{{\bf 
k}}^s$ is the energy of the coherent motion of the Zhang-Rice singlet, 
dressed by spin fluctuations. 
The minimal energy that one needs to excite such a pair is
$$
E_{\min }^{e-h}\approx \Delta -\epsilon _{\min }^s-\left( -4\tau -\epsilon
_{\min }^{t-J}\right) \sim \Delta -2.2\tau \approx 2.6\ {\rm eV} \; . 
$$
Here we have taken into account that the minimum of the spectrum in the 
$t-J$ model
is $\epsilon _{\min }^{t-J}\sim -2\tau $ and that the Zhang-Rice singlet
energy is $\epsilon _{\min }^s\sim -4.2\tau $. \cite{Kuz98} This result 
indicates that for
layered cuprates the excitonic feature is immersed into the electron-hole
continuum and represents a resonance rather than a discrete level. Of course,
for the the final conclusion more detailed calculations should be performed 
within an enlarged basis set.

Summarizing the calculated spectra we may state a rough, qualitative,
agreement with the experimental curves \cite{Wang96,Fink01} already within the
'minimal' version of the Emery model and using a minimal basis set (Figs.\
1-3). One may note a remarkable
influence of the background dielectric function $\varepsilon_{\infty}$ (Figs.\
2 and 3). We 
found that additional parameters like e.g.\ $t_{pp}$ or $U_p$ act
in different ways. That might improve a future parameter fit of
the experimental curves. But the present accuracy is not sufficient for a
reliable fit, which has to be reserved for the future.

In this work we have not taken into account the intersite Coulomb repulsion 
which leads to electron hole attraction. This term has a twofold influence on
the position of the excitonic feature. From one hand, it leads to an effective
increase of the fundamental gap, from the other hand it contributes to the
electron-hole binding. These tendencies are opposite to each other and should
be thoroughly explored in a separate investigation.

\section*{Acknowledgments}

This work was supported by DFG (436 UKR 113/49/41) and by RFFR (project N
01-12-16719). R.O.K. and A.F.B. thank for hospitality the IFW Dresden, where
the main part of this work has been carried out. We wish to thank S.-L.
Drechsler and J. M\'alek for numerous and useful discussions.

\appendix*

\section*{Appendix}

Here we give the formulae for the matrices (\ref{SandL}) for the Hamiltonian (
\ref{Htot}) that includes also a finite oxygen on-site repulsion $U_p$ and
a direct O-O hopping $t_{pp}$. The effective Hamiltonian may be generally
derived by the following procedure. Let be 
\begin{equation}
\label{H}\hat H=\hat H_0+\hat V,\ \hat H_0=\sum_m\left| m\right\rangle
E_m\left\langle m\right| ,\ \hat V=\sum_m\left| n\right\rangle
t_{nm}\left\langle m\right| , 
\end{equation}
then the canonical transformation with the operator 
\begin{equation}
\label{SW}\hat S=\sum_m\left| n\right\rangle \frac{t_{nm}}{E_m-E_n}
\left\langle m\right| 
\end{equation}
gives $\left[ \hat H_0,\hat S\right] =-\hat V$, and finally up to the second
order
$$
\hat H_{eff}=\exp (-\hat S)\hat H\exp (\hat S)=H_0+\frac 12\left[ \hat V,
\hat S\right] = 
$$
\begin{equation}
\label{Heffg}=\frac 12\sum_m\left| n\right\rangle \left( \frac{t_{nj}t_{jm}}{
E_m-E_j}+\frac{t_{nj}t_{jm}}{E_n-E_j}\right) \left\langle m\right| . 
\end{equation}

Within the operator basis containing (\ref{psi}) and (\ref{xi}) we have the
equations of motion for the Hamiltonian containing finite $U_p,t_{pp}$
$$
[\psi _{{\bf R,}\alpha },\hat H]=E_0\psi _{{\bf R,}\alpha }+\tau \sum_\beta
\psi _{{\bf R,}\beta }+\tau _u\xi _{{\bf R},\alpha ,-\alpha }+[\tau _u\xi _{
{\bf R}+{\bf g}_\alpha ,-\alpha ,\alpha }-(\tau _u-\tau )\psi _{{\bf R}+{\bf 
g}_\alpha ,-\alpha }]-t_{pp}\sum_{\beta \neq \alpha ,-\alpha }\psi _{{\bf R,}
\beta } 
$$
$$
[\xi _{{\bf R},\alpha ,-\alpha },\hat H]=E_0\xi _{{\bf R},\alpha ,-\alpha
}+\tau \xi _{{\bf R}\alpha ,-\alpha }+\tau \sum_{\beta \neq \alpha }Z_{{\bf R
}+{\bf g}_\alpha }^{\gamma _1,\gamma _2}Z_{{\bf R}}^{\gamma _2,0}p_{{\bf R}+
{\bf a}_\beta ,\gamma _1}-t_{pp}\sum_{\beta \neq \alpha ,-\alpha }Z_{{\bf R}+
{\bf g}_\alpha }^{\gamma _1,\gamma _2}Z_{{\bf R}}^{\gamma _2,0}p_{{\bf R}+
{\bf a}_\beta ,\gamma _1}+ 
$$
$$
+\tau _u\psi _{{\bf R},\alpha }+\tau _u\psi _{{\bf R}+{\bf g}_\alpha
,-\alpha }-(\tau _u-\tau )\xi _{{\bf R}+{\bf g}_\alpha ,-\alpha ,\alpha }. 
$$
Here $\tau _u\equiv t^2/(\Delta +U_p)$ and $E_0=\Delta +4\tau +2(\tau -\tau
_u)$ .
It is convenient to introduce the notation 
$$
\omega _{{\bf r}}\equiv \sum_{\gamma ,\gamma _1,\gamma _2,\ldots }\langle
Z_0^{\gamma ,\gamma _1}Z_{{\bf g}}^{\gamma _1,\gamma _2}\ldots Z_{{\bf r}
}^{\gamma _r,\gamma }\rangle . 
$$
For ${\bf r}$ up to the third neighbors $\omega _{{\bf r}}$ is expressed via
two point correlation functions 
\begin{equation}
\omega _g=\frac 12+\left\langle {\bf \hat S}_{{\bf R}}\cdot {\bf \hat S}_{
{\bf R+g}}\right\rangle ,\ \omega _{{\bf g}_\alpha +{\bf g}_\beta }=\frac 14
+2\left\langle {\bf \hat S}_{{\bf R}}\cdot {\bf \hat S}_{{\bf R+g}
}\right\rangle +\left\langle {\bf \hat S}_{{\bf R}}\cdot {\bf \hat S}_{{\bf 
R+g}_\alpha +{\bf g}_\beta }\right\rangle \; .
\end{equation}
The Liouvillean and overlap matrices in ${\bf k}$-space are then  
\begin{equation}
\label{Sk}\langle \{\xi _{{\bf k}\alpha ,-\alpha },\psi _{{\bf k}\beta
}^{\dagger }\}\rangle \;=\;\delta _{\alpha \beta }\omega _g \; ,
\end{equation}
\begin{equation}
\label{Hk}\langle \{[\psi _{{\bf k}\alpha },\hat H],\psi _{{\bf k}\beta
}^{\dagger }\}\rangle \;=\delta _{\alpha \beta }\left( \tau +\tau _u\omega
_g\right) +\left( \tau -t_{pp}\right) \left( 1-\delta _{\alpha \beta
}\right) \left( 1-\delta _{\alpha ,-\beta }\right)  
\end{equation}
\begin{equation}
+\delta _{\alpha ,-\beta }\left[ \tau +\exp (\imath {\bf kg_\alpha })\left(
\tau _u\omega _g-\tau _u+\tau \right) \right] \; ,
\end{equation}
\newpage
\begin{equation}
\langle \{[\xi _{{\bf k}\alpha ,-\alpha },\hat H],\psi _{{\bf k}\beta
}^{\dagger }\}\rangle \;=\delta _{\alpha \beta }\left( \tau _u+\tau \omega
_g\right) +\left( \tau -t_{pp}\right) \omega _g\left( 1-\delta _{\alpha
\beta }\right) \left( 1-\delta _{\alpha ,-\beta }\right)  
\end{equation}
$$
+ \delta _{\alpha ,-\beta }\left\{ \tau \omega _g+\exp (\imath {\bf kg_\alpha }
)[\tau _u-(\tau _u-\tau )\omega _g]\right\} \; ,
$$
\begin{equation}
\langle \{[\xi _{{\bf k}\alpha ,-\alpha },\hat H],\xi _{{\bf k}\beta ,-\beta
}^{\dagger }\}\rangle \;=[\delta _{\alpha \beta }(\tau +\tau _u\omega
_g)+\left( 1-\delta _{\alpha \beta }\right) \left( 1-\delta _{\alpha ,-\beta
}\right) \left( \tau -t_{pp}\right) \omega _{{\bf g}_\beta +{\bf g}_\alpha
}] 
\end{equation}
$$
+ \delta _{\alpha ,-\beta }\left\{ \tau \omega _{2g}+\exp (\imath {\bf 
kg_\alpha })[\tau _u\omega _g-(\tau _u-\tau )]\right\} \; .  
$$



\begin{figure}[htb]
\label{fig1}
\epsfig{file=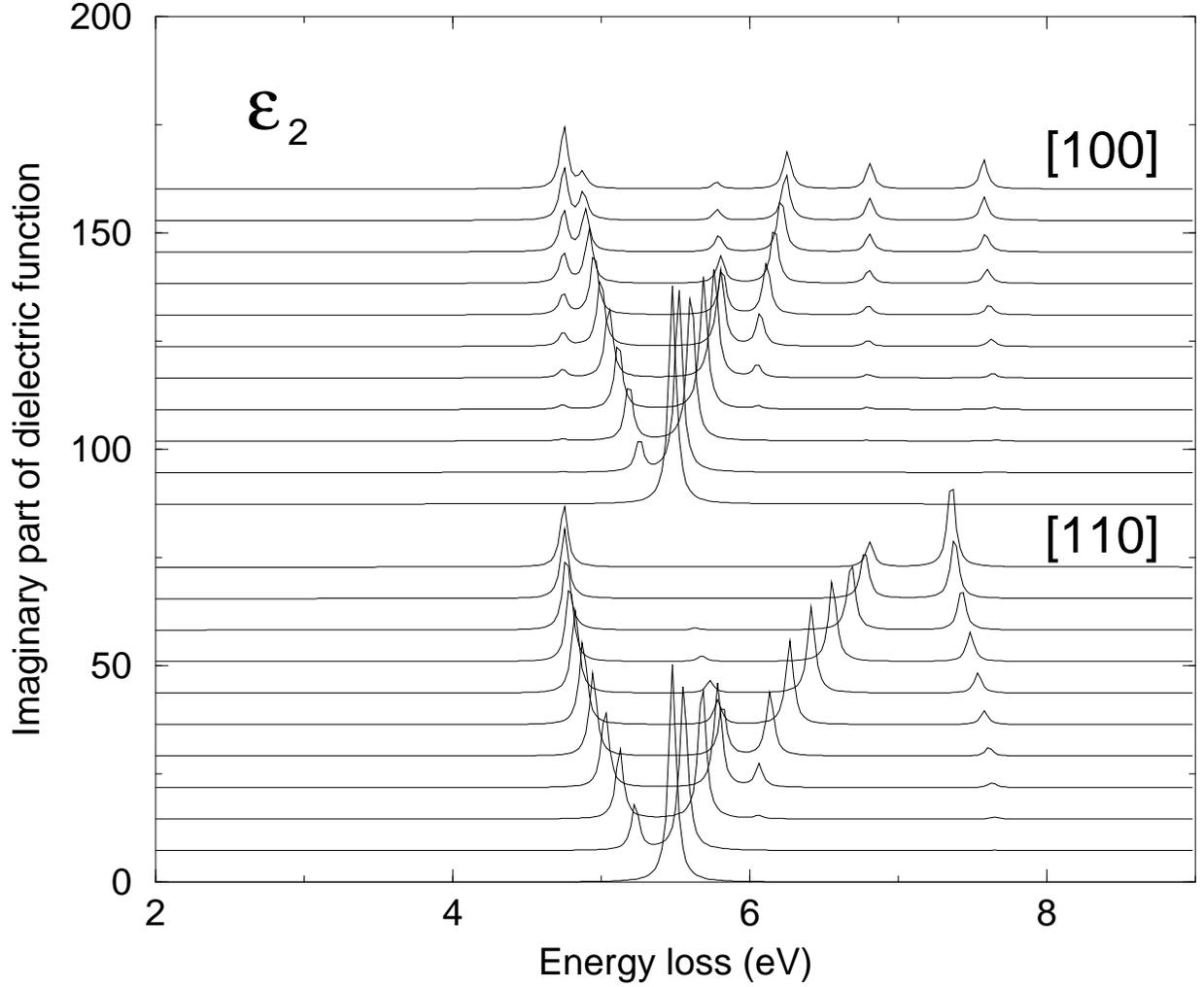}
\caption{
The imaginary part of dielectric function 
$\varepsilon _2({\bf q},\omega )\equiv {\rm Im}\varepsilon ({\bf q},\omega )$
as a function of frequency $\omega $ and wave vector ${\bf q}$ along 
two symmetry directions. For each direction the curve number $n$ 
corresponds to 
$q_x = \pi n/10a $, $a$ being the lattice constant, $n=0$ for the bottom curve.
The parameters are $\Delta =3.6\ $eV and $t=1.3$ eV, 
oxygen on-site repulsion $U_p$ and O-O hopping $t_{pp}$ were neglected.}
\end{figure}
\newpage

\begin{figure}[htb]
\label{fig2a}
\epsfig{file=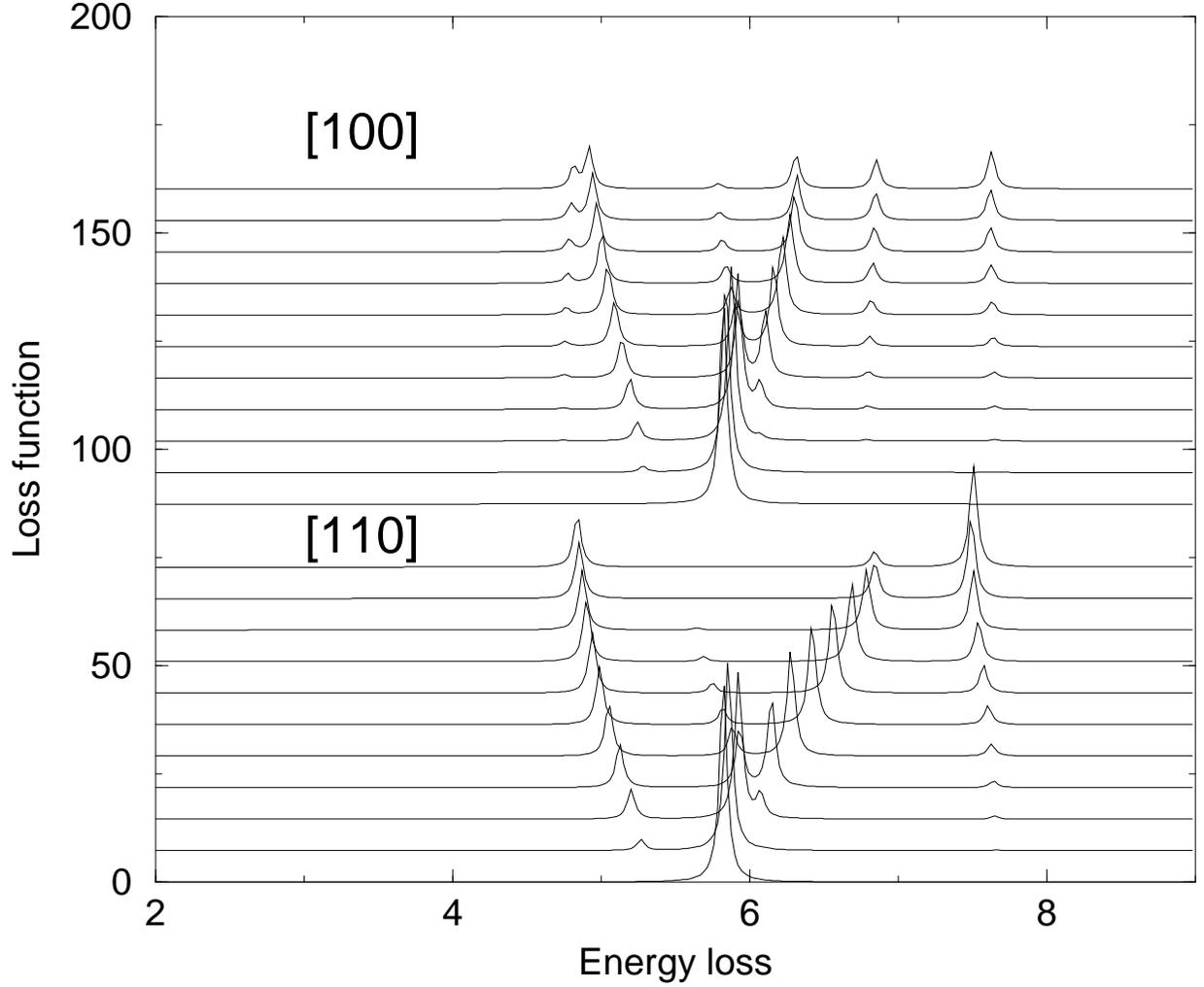}
\caption{
The loss function 
$L({\bf q},\omega )\equiv -{\rm Im}\left[ \varepsilon ^{-1}({\bf q}
,\omega )\right]$  (multiplied by $\left[ \varepsilon _\infty ({\bf 0}
)\right] ^2$ in order to have aproximately the same normalization as $
\varepsilon _2({\bf q},\omega )$)  for the same parameters  as in Fig.\ 1  and
under the assumption
that $\varepsilon ({\bf q},\omega =0)\approx \varepsilon ({\bf q=0}
,\omega =0)=4.83$. }
\end{figure}

\newpage
\begin{figure}[htb]
\label{fig3}
\epsfig{file=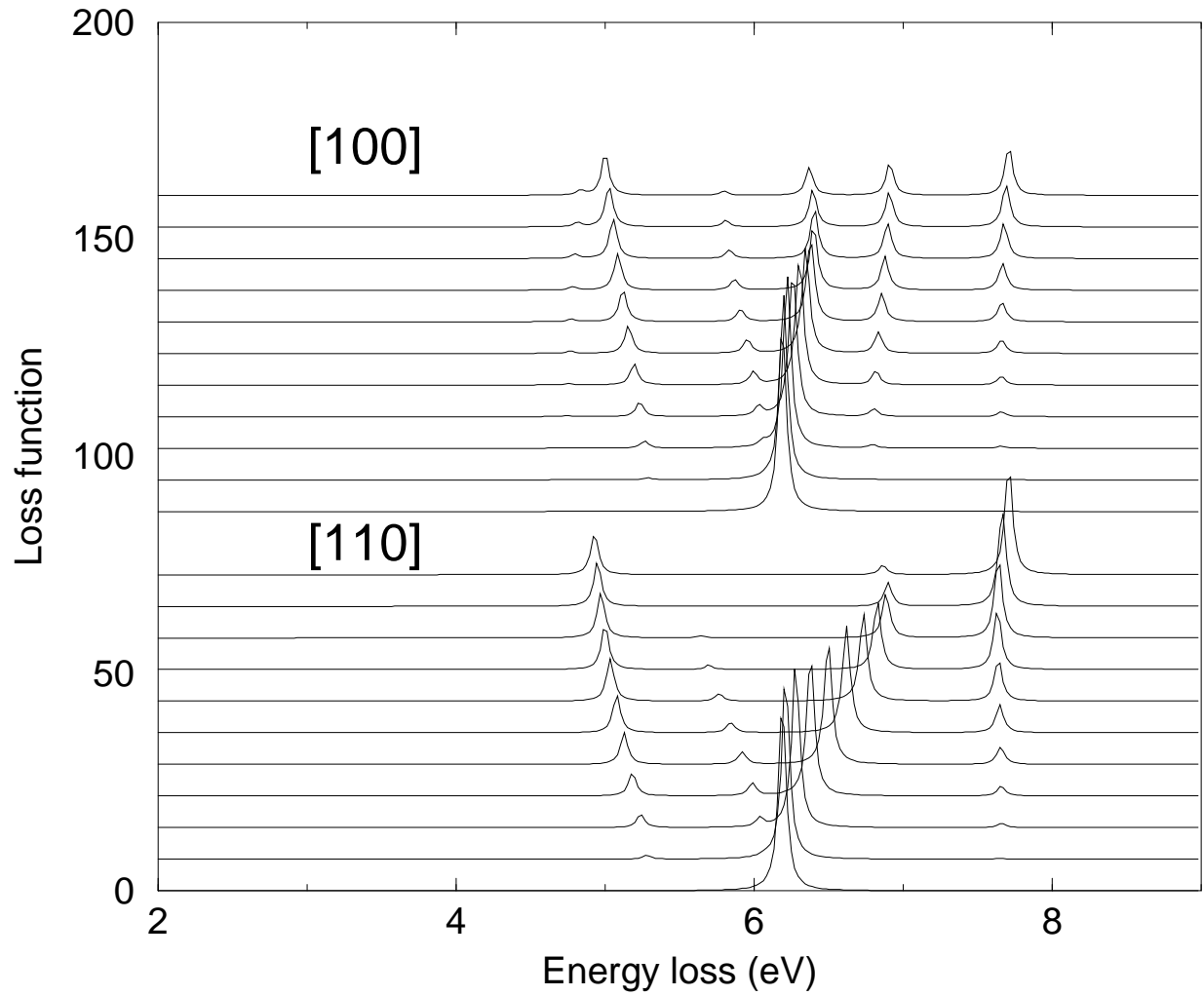}
\caption{
The loss function for 
constant $\varepsilon _\infty =2$ for all ${\bf q}$ and the parameter 
set of Fig.\ 1. }
\end{figure}

\newpage

\begin{figure}[htb]
\label{fig4}
\epsfig{file=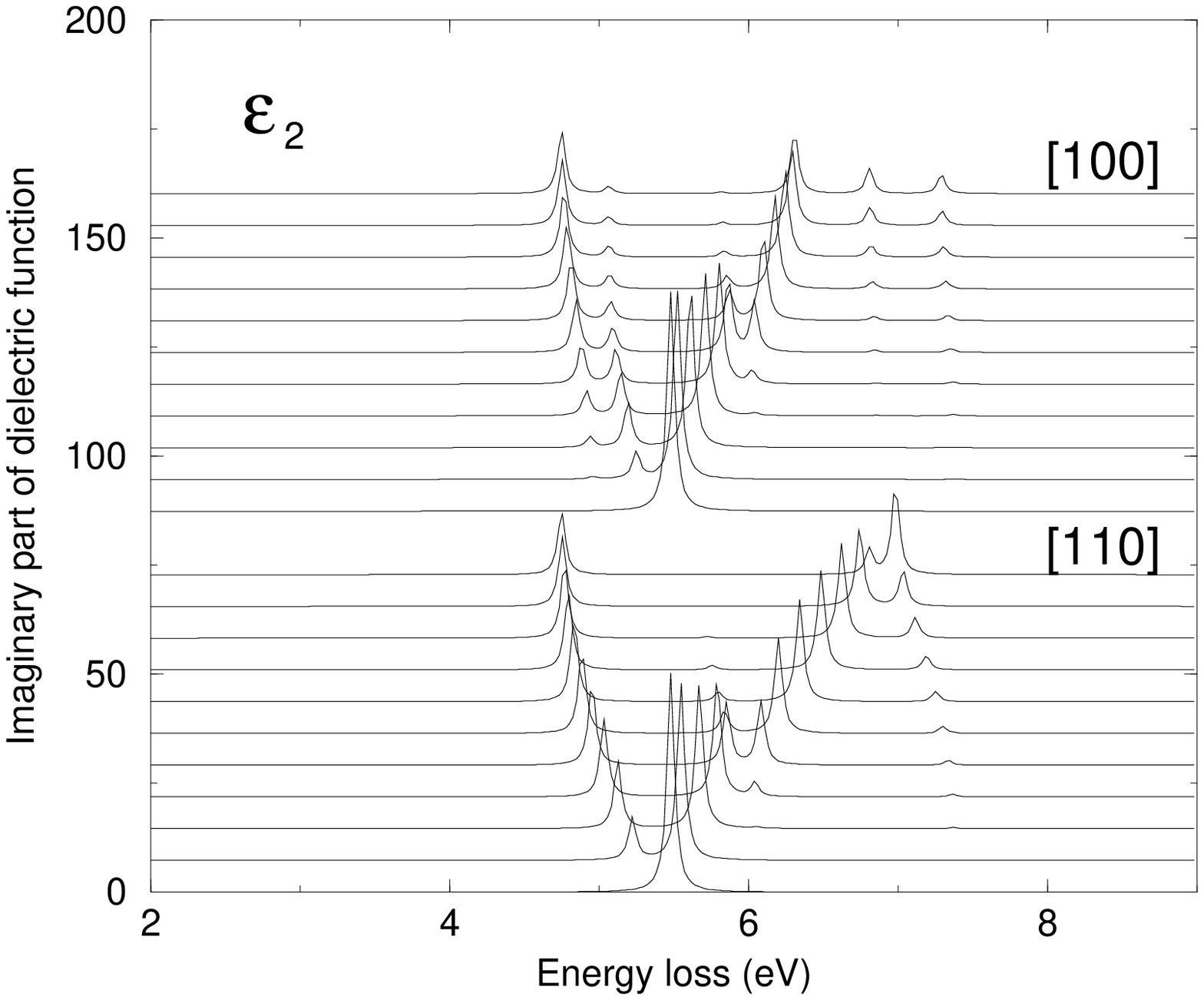}
\caption{
 $\varepsilon _2({\bf q},\omega )$ for $
t_{pp}=0.4\tau $ and $U_p=0$. The other parameters are the same as in Fig.\ 1.}
\end{figure}

\newpage

\begin{figure}[htb]
\label{fig5}
\epsfig{file=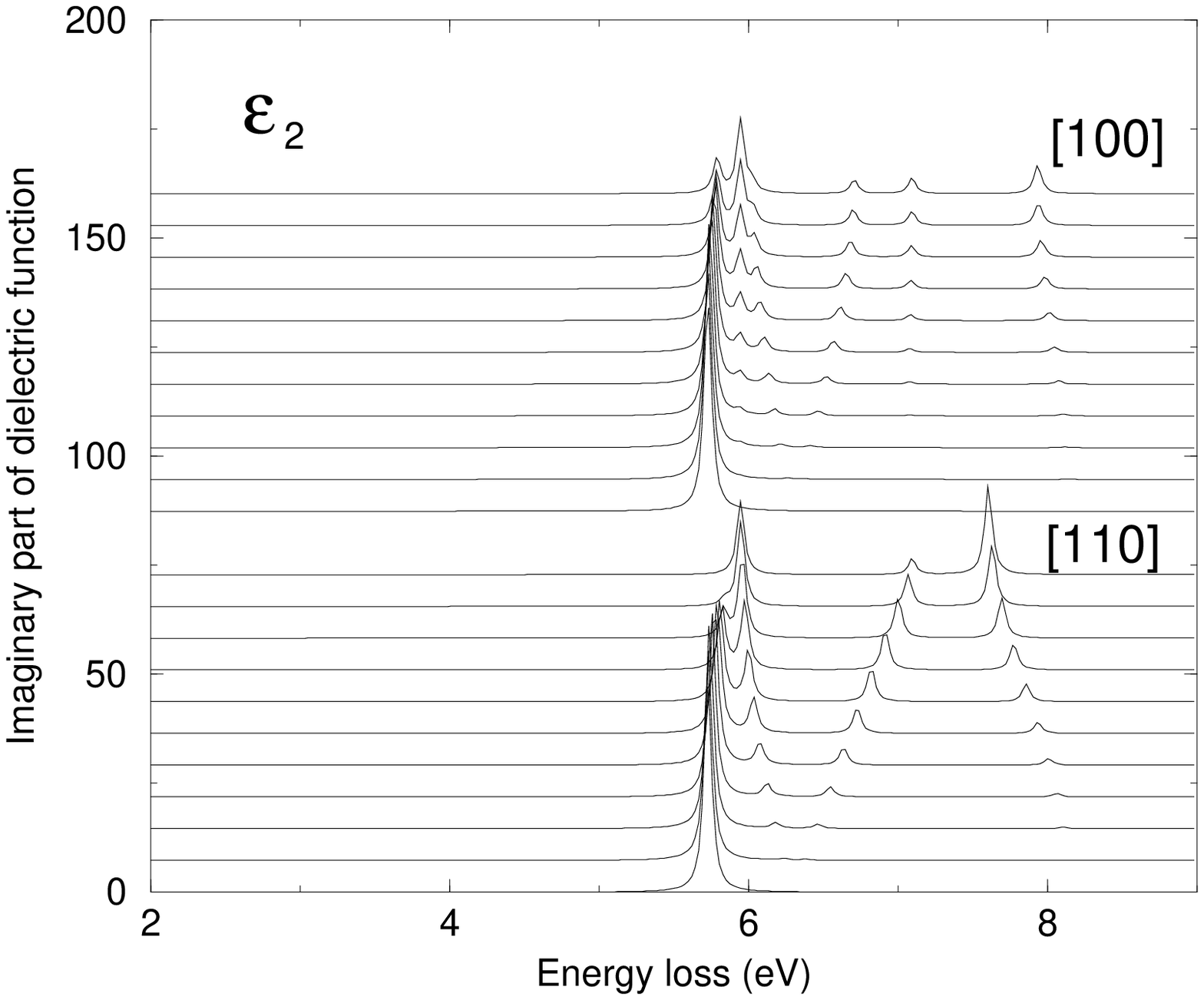}
\caption{
 $\varepsilon _2({\bf q},\omega )$ for $
t_{pp}=0$ and $U_p=4 eV$. The other parameters are the same as in Fig.\ 1. }
\end{figure}


\end{document}